\documentclass[pra,aps,showpacs,floatfix,twocolumn]{revtex4}
\usepackage{graphicx}
\usepackage{array}
\usepackage{color}
\usepackage{amsmath}
\usepackage{amsxtra}
\usepackage{amstext}
\usepackage{amssymb}
\usepackage{latexsym}
\usepackage{dsfont}

\usepackage[colorlinks,
            linkcolor=blue,
            anchorcolor=blue,
            citecolor=blue,
            urlcolor=blue
            ]{hyperref}

\begin{document}

\title{Resonant synchronization and information
retrieve from memorized Kuramoto network}
\author{Lin Zhang}
\email[Email address: ]{zhanglincn@snnu.edu.cn}
\author{Xv Li}
\author{Tingting Xue}
\affiliation{School of Physics and Information Technology, Shaanxi Normal University,
Xi'an 710061, P. R. China}

\begin{abstract}
A new collective behavior of resonant synchronization is discovered and the
ability to retrieve information from brain memory is proposed based on this
mechanism. We use modified Kuramoto phase oscillator to simulate the dynamics
of a single neuron in self-oscillation state, and investigate the collective
responses of a neural network, which is composed of $N$ globally coupled Kuramoto
oscillators, to the external stimulus signals in a critical state just below
the synchronization threshold of Kuramoto model. The input signals at different
driving frequencies, which are used to denote different neural stimuli, can
drive the coupled oscillators into different synchronized groups locked to
the same effective frequencies and recover synchronized patterns emerged from their collective
dynamics closely related to the predetermined frequency distributions of the oscillators (memory).
This model is used to explain how brain stores and retrieves information from the synchronization patterns
emerging in the neural network stimulated by the external inputs.
\end{abstract}

\date{\today }
\maketitle
\section{Introduction}
The consciousness of the human brain and its memory mechanism are the
most fascinating problems in physics and biology \cite{A26,A27,A28}. With
the help of molecular biology and the brain magnetic resonance imaging
(MRI), neurobiologists have identified the detailed physical structures and
the local functions of the brain \cite{A29}. Regions of the brain have
been proved to be a complex network of well-connected neurons working in a
critical state that can exhibit complicated synchronization behaviors during the
brain's continuous discharge induced by neutral stimulus from our
sensory organs \cite{A17}. As synchronized patterns of neurons are indicated
to demonstrate the cognitive activities in human brain, many
mathematical models have been proposed to study consciousness activities in
human brain \cite{A46}. However, due to the complexity of the neural network,
analytical analysis cannot be carried out efficiently on theses models and most
numerical works are limited to give apparent behaviors of cognitive activity \cite{A47}.
Based on the synchronized spatiotemporal patterns emerged in brain
activities \cite{A42}, we adopt Kuramoto model \cite{A1,A2}, a very simple model
of globally coupled nonlinear rotators, to study the collective dynamics of neutrons working
in a state of self-sustained oscillation and explore their collective behaviors by resonant
synchronization phenomenon. Our study shows that Kuramoto model presents a
synchronized resonant behavior and which can be used to study
presentation or extraction of information from human brain by external
stimulation \cite{A19,A33,A30,A31,A32}. By studying the collective
resonance of Kuramoto model, we find that when the coupling strength between
Kuramoto oscillators (KOs) is just below the synchronization threshold, the coupled
KOs will exhibit an obvious resonant synchronization when the system
is subjected to a resonant external stimulus, and it can exactly provide an
explanation of the brain memory to store and retrieve information in
its critical neural networks.
\section{The resonant synchronization}

\subsection{The synchronized dynamics between two coupled harmonic oscillators by external driving}
\begin{figure}[tph]
\includegraphics[width=0.45\textwidth]{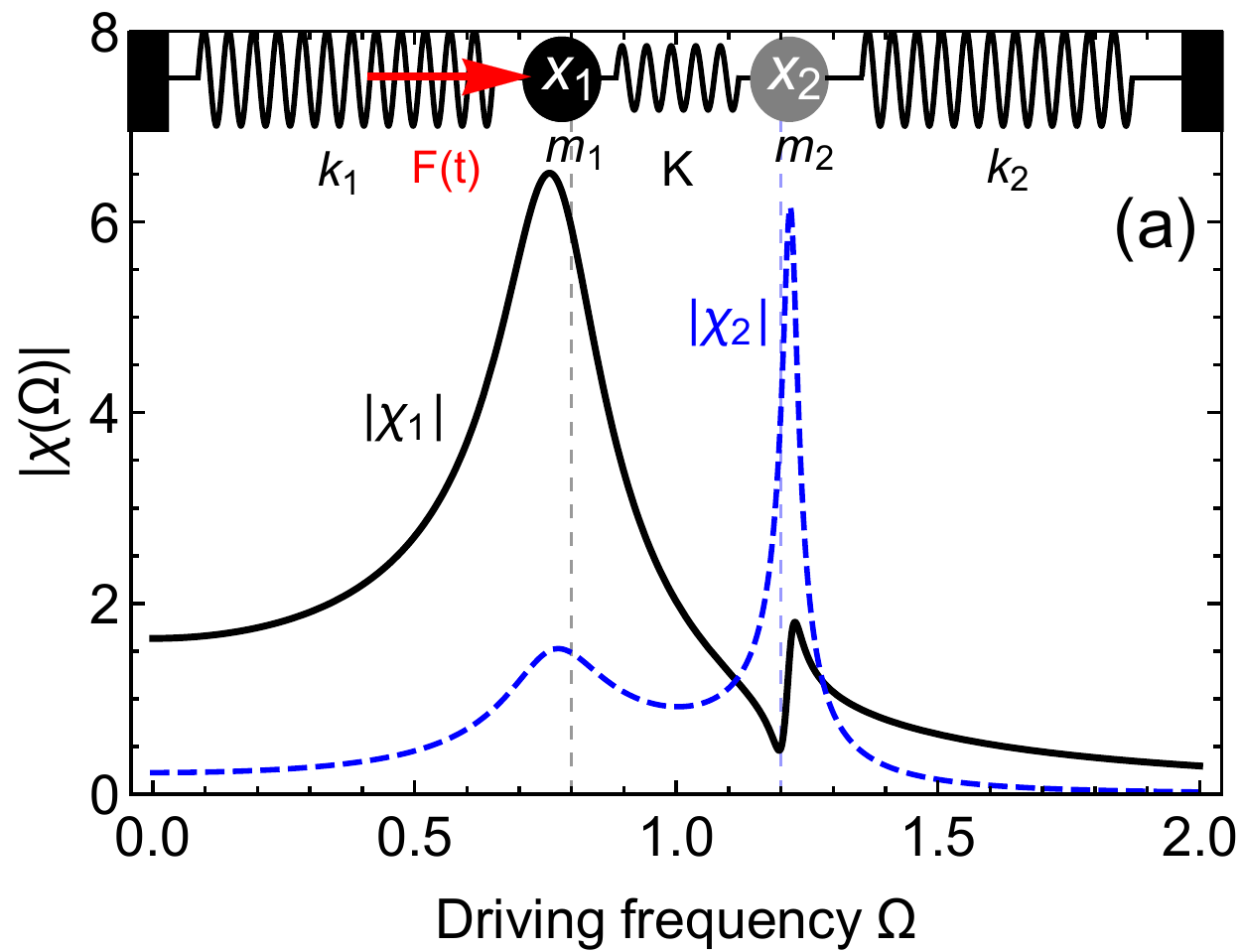}\newline\\
\includegraphics[width=0.45\textwidth]{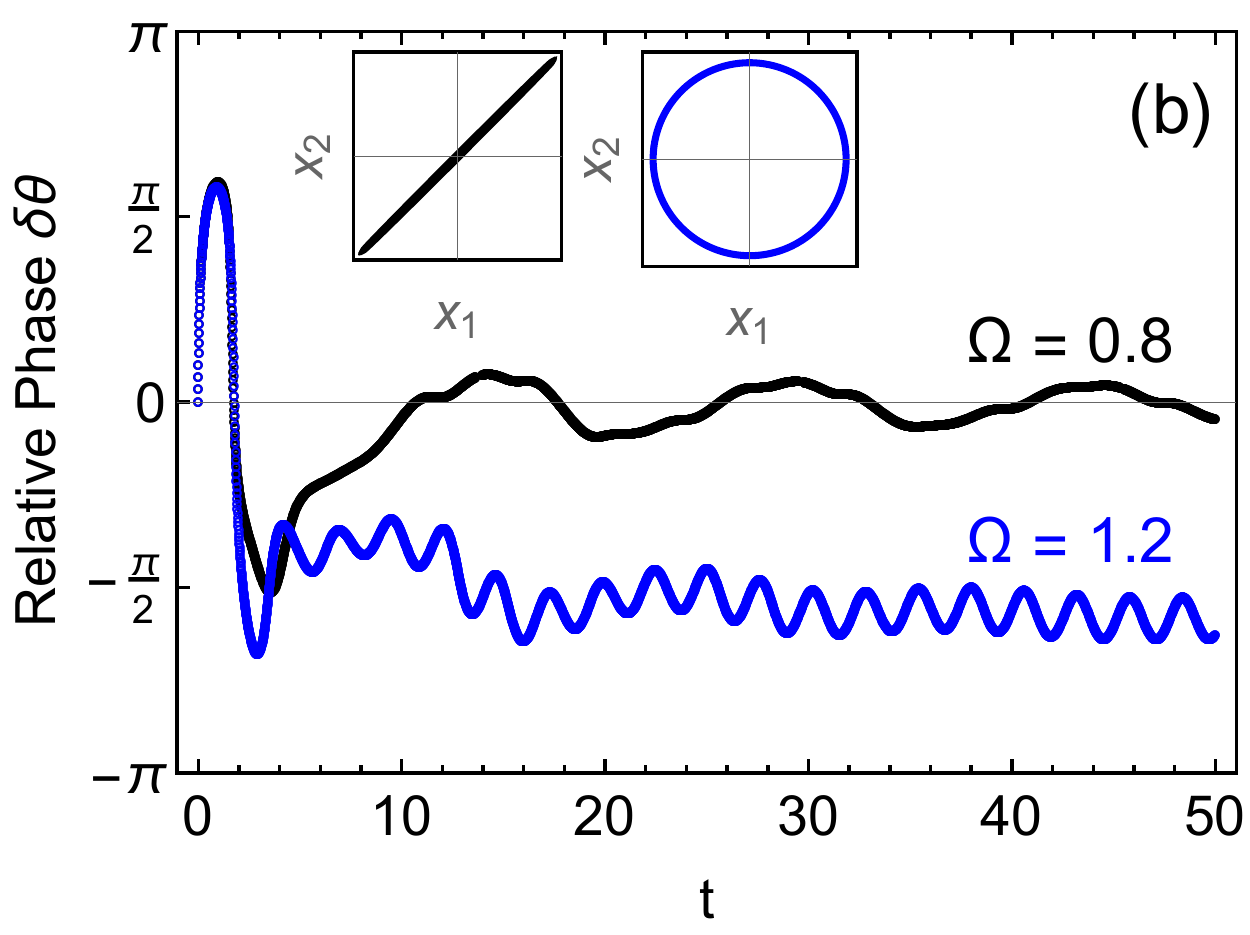}\newline
\caption{(color online). (a) The amplitude responsive spectra of two coupled HOs
driven by the external force on the first oscillator
labeled by $x_{1}$. The vertical dashed lines indicate the driving frequencies of
$\Omega=\omega _{R1}\approx 0.8$ and $\Omega=\omega _{R2}\approx 1.2$. Top inset:
the configuration of the two coupled HOs driven by force $F(t)$.
(b) The relative phase of two
HOs driven by resonant frequencies of $\Omega =\omega _{R1}$
and $\Omega =\omega _{R2}$, respectively. Top inset: the longtime related obits
between $x_1(t)$ and $x_2(t)$ at driving frequencies of $\Omega=0.8$ (left) and
$\Omega=1.2$ (right), respectively. The other parameters are $\protect\gamma _{1}=0.1,%
\protect\gamma _{2}=0.01$ and $K_1=K_2=0.2$. }
\label{figure2}
\end{figure}

In order to give a basic idea of the resonant synchronization, we begin with a well-studied model:
two coupled linear harmonic oscillators (HOs). First, resonance is a
very common dynamic phenomenon in physics, in which a vibrating system or an
external force drives another system to oscillate with an increased
energy at certain frequencies \cite{A36}. Mechanical resonance can accumulate energy to
break glass cups or even damage bridges and
buildings \cite{A13,A34,A35,A14,A25}. The dynamical equation to describe resonance
is
\begin{equation}
\ddot{x}+2\gamma \dot{x}+\omega _{0}^{2}x=F_{0}\cos \Omega t,  \label{FDh}
\end{equation}%
where $x$ describes the vibration displacement, $\omega _{0}$ is the
eigenfrequency of the oscillator, $\gamma$ is the damping rate, $F_{0}$
and $\Omega$ are the driving strength and frequency, respectively. The
motion of the driven oscillator in frequency domain
reads
\begin{equation}
x\left( \omega \right) =\chi \left( \omega \right) F\left( \omega \right) ,
\label{Hr}
\end{equation}%
where $\chi \left( \omega \right) $ is a complex responsive function
\begin{equation}
\chi \left( \omega \right) =\frac{1}{\left( \omega _{0}^{2}-\omega
^{2}\right) -i\left( 2\gamma \omega \right) ,}
\end{equation}%
which is also called displacement susceptibility. Then the solution of Eq.(\ref{FDh})
is an inverse Fourier transformation of Eq.(\ref{Hr}) by
\begin{equation}
x\left( t\right) =\left\vert \chi \left( \Omega \right) \right\vert
F_{0}\cos \left( \Omega t-\theta \right) ,
\end{equation}%
where $\theta =\arg \chi \left( \Omega \right) $ is the phase
of $\chi \left( \Omega \right) $. Obviously, after a transient dynamics, the
frequency of the driven oscillator is locked to driving frequency $%
\Omega $ and its vibration amplitude (energy) reaches maximum when
$\Omega \rightarrow \omega _{R}=\sqrt{\omega _{0}^{2}-2\gamma ^{2}}$, which
is the so called resonant phenomenon.
For a high-$Q$ oscillator $\gamma \rightarrow 0$, an obvious dynamical
transition between in-phase and out-of-phase motions demonstrates an important
synchronized behavior at driving point of $\omega_{R}\approx \omega _{0}$. This behavior,
especially happens among an ensemble of coupled nonlinear oscillators in a complicated networks, is
called resonant synchronization: synchronization
induced by resonant driving. In order to illustrate this, we return to
two coupled HOs in a configuration of Fig.\ref{figure2}(a) (top inset) which are described by
\begin{eqnarray}
\ddot{x}_{1}\left( t\right) +2\gamma _{1}\dot{x}_{1}\left( t\right) +\omega
_{01}^{2}x_{1}\left( t\right)  &=&K_{1}x_{2}\left( t\right) +F(t), \\
\ddot{x}_{2}\left( t\right) +2\gamma _{2}\dot{x}_{2}\left( t\right) +\omega
_{02}^{2}x_{2}\left( t\right)  &=&K_{2}x_{1}\left( t\right) ,
\end{eqnarray}%
where\begin{eqnarray}
\omega_{01}=\sqrt{\frac{k_{1}+K}{m_1}}, K_1=\frac{K}{m_1},\\
\omega_{02}=\sqrt{\frac{k_{2}+K}{m_2}}, K_2=\frac{K}{m_2},
\end{eqnarray}%
with $k_1$, $k_2$ and $K$ being the stiffness coefficients of the springs labeled in
Fig.\ref{figure2}(a). Then the displacement responses of two coupled HOs are
\begin{eqnarray}
x_{1}\left( \omega \right)  &=&\chi _{1}\left( \omega \right) F(\omega ), \\
x_{2}\left( \omega \right)  &=&\chi _{2}\left( \omega \right) F(\omega ),
\end{eqnarray}%
where the displacement susceptibilities are
\begin{eqnarray*}
\chi _{1}\left( \omega \right)  &=&\frac{\left( \omega _{02}^{2}-\omega
^{2}\right) -i2\gamma _{2}\omega }{\left( \omega _{01}^{2}-\omega
^{2}-i2\gamma _{1}\omega \right) \left( \omega _{02}^{2}-\omega
^{2}-i2\gamma _{2}\omega \right) -k_{1}k_{2}}, \\
\chi _{2}\left( \omega \right)  &=&\frac{k_{2}}{\left( \omega
_{01}^{2}-\omega ^{2}-i2\gamma _{1}\omega \right) \left( \omega
_{02}^{2}-\omega ^{2}-i2\gamma _{2}\omega \right) -k_{1}k_{2}}.
\end{eqnarray*}%
Fig.\ref{figure2}(a) shows that the amplitude responses of $|\chi _{1,2}(\Omega)|$
to force $F(t)=F_{0}\cos \Omega t$ exerted on the first
oscillator exhibits two types of resonant peak.
The synchronized properties at two resonant
frequencies around $\Omega =\omega _{R1}$ and $\Omega =\omega _{R2}$ are shown
in Fig.\ref{figure2}(b) by the relative phase between two HOs
\begin{equation}
\delta \theta =\arg (\chi _{1})-\arg (\chi _{2}).
\end{equation}%
We can see that when the driving frequency comes near to the resonant frequencies $%
\omega _{R}$, the relative phase tends to be connected due to the resonant
effect, such as the lower resonant driving frequency $\omega _{R1}$ leading to an
in-phase vibration (left top inset of Fig.\ref{figure2}(b)), and the higher resonant
frequency $\omega _{R2}$ leading
to a $\pi /2$-lagged vibration (right top inset of Fig.\ref{figure2}(b)).
This toy model demonstrates that the resonant driving can give manifest influences
on the collective amplitude and phase of coupled oscillators, which will lay down the main
idea of resonant synchronization for information store and retrieve from the network.

\subsection{Resonant synchronization in Kuramoto model}

Basically, synchronization is the correlated dynamical behaviors between coupled nonlinear
oscillators, but, for linear HOs, the above correlated dynamics shown in Fig.\ref{figure2} is not strictly
synchronization \cite{A16}. In 1673, Huygens observed synchronized motion
of two adjacent pendulum clocks and opened the study of synchronization,
and then various synchronizations are found to reveal a universal
phenomenon in nature. For nonlinear oscillators, synchronization is related to the
self-sustained (limit-cycle) oscillation, which is a special state that has stable amplitude and
free phase \cite{A15}. Roughly, there are two important types of synchronization, one is the driven
synchronization induced by a stable driving, and the other is the mutual synchronization
due to the mutual couplings between oscillators. What we study here is the driven
type whose behaviors are closely dependent on the coupling strength and
the driving detuning \cite{A16}. In recent years, the synchronized behaviors of
neurons have received widespread attentions in the study of vision \cite{A3},
movement \cite{A4}, memory \cite{A5}, epilepsy \cite{A6,A7,A8,A9,A10,A11,A12} and so on.

The Kuramoto model is a typical physical model to exhibit synchronization when the coupling
rate $k$ reaches a critical value $k_c$.
In a system with only two coupled KOs, the nonlinear equations
of the system are
\begin{eqnarray}
\dot{\theta}_{1} &=&\omega _{1}+\frac{k}{2}\sin \left( \theta _{2}-\theta
_{1}\right), \\
\dot{\theta}_{2} &=&\omega _{2}+\frac{k}{2}\sin \left( \theta _{1}-\theta
_{2}\right),
\end{eqnarray}
where each oscillator has its own eigenfrequency $\omega _{1}$ or $\omega _{2}$, and $k$ is their
coupling rate. The system can describe two neutrons in limit-cycle states with their pulsing phases
$\theta_{i}$. There are two ways to synchronize KOs:
increasing coupling strength $k$ and introducing external driving. In this paper,
the driving is introduced to simulate the imagination or thinking process triggered by the neural signals
from our sensory organs.
Then the full equations are \cite{A43}
\begin{eqnarray}
\dot{\theta_{1}}=\omega_{1}+\frac{k}{2}\sin(\theta_{2}-\theta_{1})
+\Lambda\sin(\Omega t-\theta_{1}),
\end{eqnarray}
\begin{eqnarray}
\dot{\theta_{2}}=\omega_{2}+\frac{k}{2}\sin(\theta_{1}-\theta_{2}),
\end{eqnarray}
where $\Lambda$ and $\Omega$  are the stimulating strength and frequency, respectively.
Keeping the coupling rate $k$ below the synchronization threshold $k_c$, that is, the
system is initially in a non-synchronous state, the KOs
synchronize after a transient state by adding an
external signal with a frequency close to the eigenfrequencies of the KOs.
\begin{figure}[tph]
\includegraphics[width=0.45\textwidth]{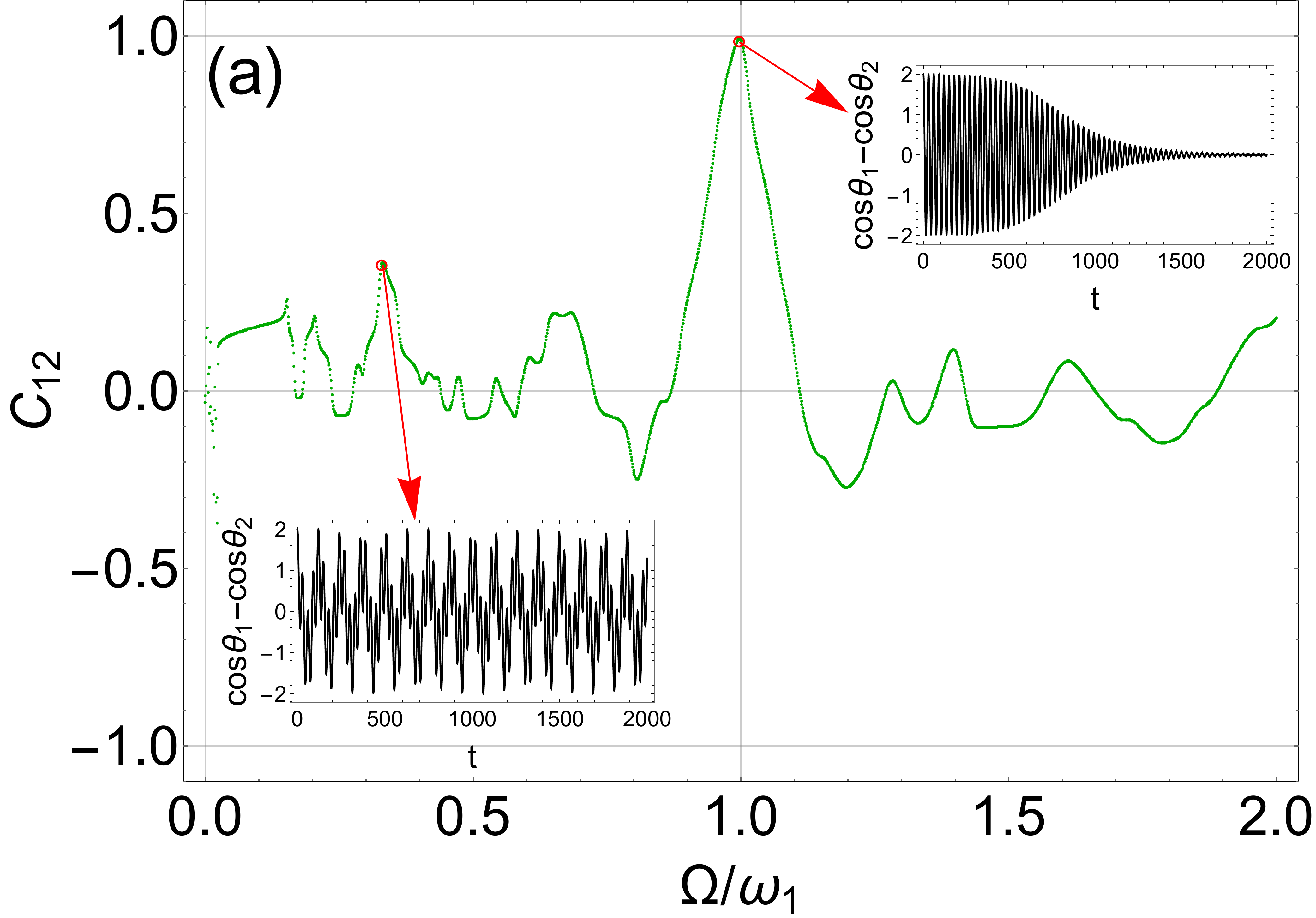}\newline
\includegraphics[width=0.45\textwidth]{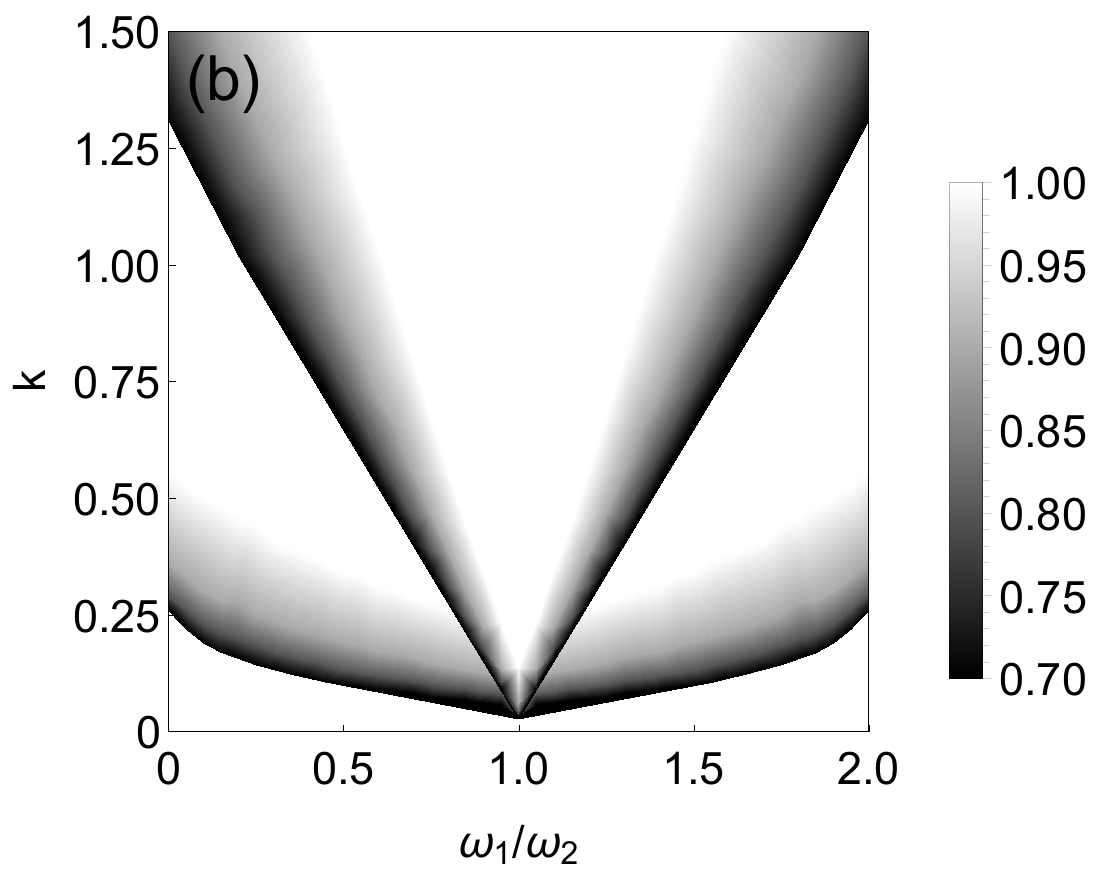}%
\caption{(color online) (a) The synchronization degree estimated by $C_{12}$ varies with respect to
the external driving frequency $\Omega$. The eigenfrequencies of two oscillators are
$\omega_{1}=0.2, \omega_{2}=0.21, $ and the coupling strength $k=0.008$. Inset: the relative
phases of two oscillators estimated by $\cos\theta_1-\cos\theta_2$ sampled at driving frequencies of
$\Omega/\omega_{1}=0.35$ and $\Omega/\omega_{1}=1$.
(b) Arnold tongue describing the synchronization region estimated by $C_{12}$. The upper tongue area
represents the synchronization without external driving, and the lower is for the case by adding the
external driving with $\Lambda=1$ and $\Omega=0.2$.}
\label{FIG2}
\end{figure}
Fig.\ref{FIG2}(a) shows the resonant synchronization through Pearson correlation coefficient $C_{12}$,
a dynamical measure to estimate the degree of synchronization between oscillators $1$ and $2$. When two KOs are completely
synchronized or anti-synchronized, the coefficient $C$ will be $1$ or $-1$~\cite{A41}.
It can be clearly seen from Fig.\ref{FIG2}(a) that when the driving frequency is near
to their eigenfrequencies, the system achieve synchronization under $k<k_c$. In order to
show this resonant synchronization effect, the Arnold tongues of synchronization area are calculated with respect to
the coupling rate $k$ and the eigenfrequency ratio $\omega_1/\omega_2$. We can see that the critical coupling
$k_c$ for resonant synchronization is clearly reduced and the area of synchronous region is significantly increased by the resonant driving.

\subsection{Resonant synchronization and synchronized groups selected by resonant driving}

Now, we applied the above resonant synchronization behavior to the information retrieve from our brain.
The nervous system of our human brain is composed of nearly $86$ billion of mutually coupled neurons \cite{A48},
where dendrites or axons of neurons connect with each other through synapses to form a high-degree
connected neural network \cite{A44}. By passing electrochemical signals through synapses between neurons,
the neurons can be excited or inhibited which makes
the discharge of neurons in the nervous system is exactly a discontinuous non-linear process \cite{A24}.
The human brain is a very complicated system \cite{A45}, and thus it is impossible
to reveal its functionality by just resolving the molecular structure and the function of individual neurons,
but should include the discharge dynamics of the whole network formed by the individual
neurons. Meanwhile, the importance of the complex network have been widely recognized in
biology \cite{A22}, sociology \cite{A23} as well as in physics \cite{A20}, and which has been used to
describe a large number of biological systems exhibiting a variety of synchronization behaviors \cite{A21}.
As the complex network firstly includes a topological structure, different connections of nodes will lead to
different collective dynamics. The simple model we used here is a global coupling Kuramoto model which can
simulate the local function of a brain region by neglecting the connection
details of the network because the collective behavior of a conscious process is somehow insensitive
to the details in a high-degree connected neural network \cite{A19}.

As the traditional Kuramoto model is very different from the environment of human brain and cannot be
applied directly, we modify the Kuramoto model to describe critical dynamics of neurons by adaptive
phase rotators \cite{A19}. The natural discharge frequency of a neuron can be represented by the
eigenfrequency of KO and the more complex frequency and irregular coupling rate between neurons
are considered by adding small harmonic modulations. When the system is under the coupling thresholds
$k<k_c$ (near to the synchronized state), we study the dynamical synchronization between
neurons when an external driving signal is added to the system in order to simulate the recall process
of brain memory. In the brain, due to the changes of synapses or the neurotransmitter diffusion,
the coupling strength between neurons will slowly change over time, and the neuron discharge
frequency will change under the external stimuli. Hence, the modified Kuramoto model will be
\begin{eqnarray}
\dot{\theta_{i}}=\omega_{i}(t)+\frac{1}{N} \sum k_{ij}(t) \sin(
\theta_{j}- \theta_{i}),
\end{eqnarray}
where $\omega_{i}(t)$ and $k_{ij}(t)$ are the
time-varying frequency of neuron $i$ and the coupling strength between $i$ and $j$ neurons, respectively,
which are supposed to be governed by the simple functions
\begin{eqnarray}
\omega_{i}(t)=\omega_{i}+\beta_{i} \sin(2 \pi f_{i} t+\varphi_{i}),
\end{eqnarray}
\begin{eqnarray}
k_{ij}(t)=k_{ij}+\mu_{ij} \sin(2 \pi g_{ij} t+\psi_{ij}),
\end{eqnarray}
where the parameters $\beta_{i}=0.1\omega_{i},$ $\mu_{ij}=0.1k_{ij},$ $f_i=1,$
$g_{ij}=0.3,$ $\varphi_{i}=\pi/2$ and $\psi_{ij}=\pi$ describe the amplitudes,
frequencies and phase offsets of the time-varying natural frequency and coupling
strength, respectively.
\begin{figure}[tbp]
\includegraphics[width=0.225\textwidth]{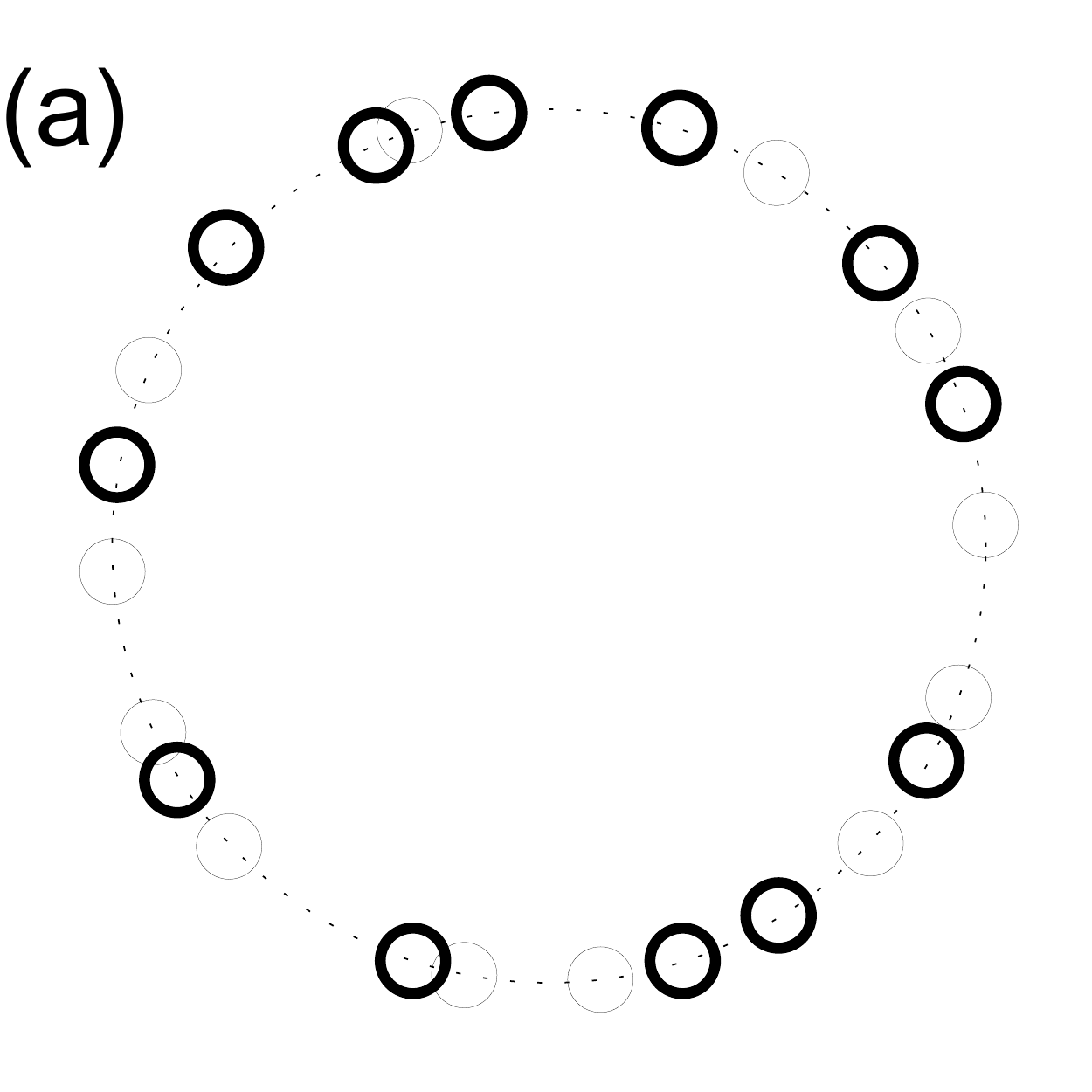}%
\includegraphics[width=0.225\textwidth]{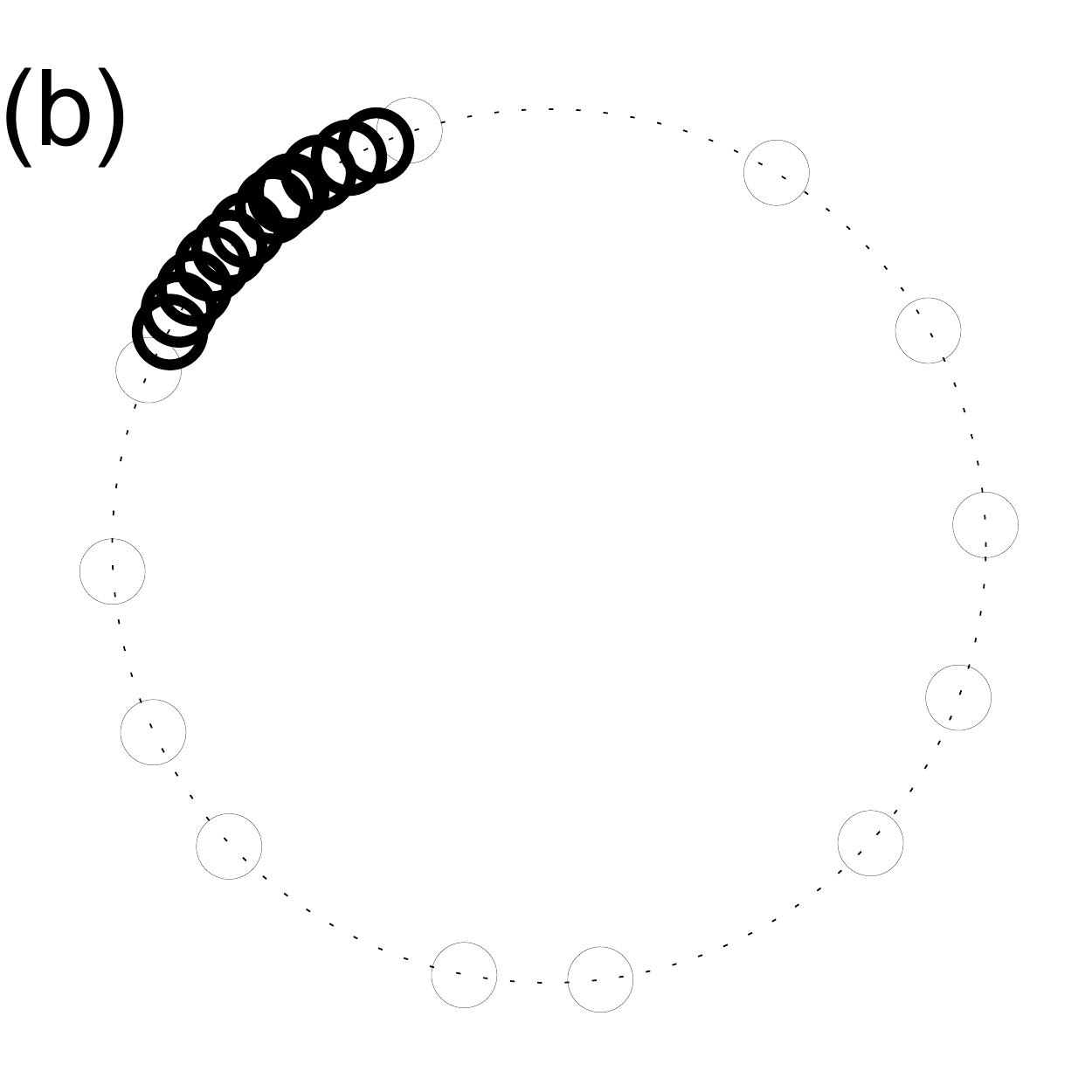}
\caption{Phase distributions of $N=12$ KOs in a phase picture
(a) without and (b) with the external driving after a same period of evolution.
The thin circles denote the initial phase distributions of KOs and the thick ones
are the phase distributions after $t=2000$.
The frequencies $\omega_i$ of the KOs are randomly sampled between the interval of $[0.9,1.1]$
and the other parameters are $\Lambda=1 , \Omega=1, k_{ij}=1$.}
\label{FIG3}
\end{figure}
In order to simulate information retrieve from the memory in a wakeful brain state
that the neuron firing is continuously activated by successive stimuli, we
introduce an activated driving of sinusoidal signal $\Lambda\sin(\Omega t-\theta_{l})$,
acting on $l^{th}$ KO (a representative oscillator selected by the stimuli in certain regions),
then the equation of motion of the $l^{th}$ oscillator becomes:
\begin{eqnarray*}
\dot{\theta_{l}}=\omega_{l}(t)+\frac{1}{N} \sum k_{lj}(t) \sin(
\theta_{j}- \theta_{l})+\Lambda\sin(\Omega t-\theta_{l}),
\end{eqnarray*}
where $\Omega$ is the characteristic frequency of external stimulus.
Fig.\ref{FIG3} simulates a globally coupled network with a number of $N=25$ KOs
and demonstrates a clear phase bunching after adding an external driving signal.
When the driving is added to one oscillator of the network, the oscillators
whose eigenfrequencies $\omega_i$ are close to the driving frequency $\Omega$
will synchronize their motions to reach a resonant synchronization state and then the
phase bunching appears. The thick circles in Fig.\ref{FIG3}(b) represents a group
of synchronized oscillators which form a synchronized
cluster, shown in the phase picture, moving together under the resonant driving input.
\begin{figure}[htp]
\includegraphics[width=0.45\textwidth]{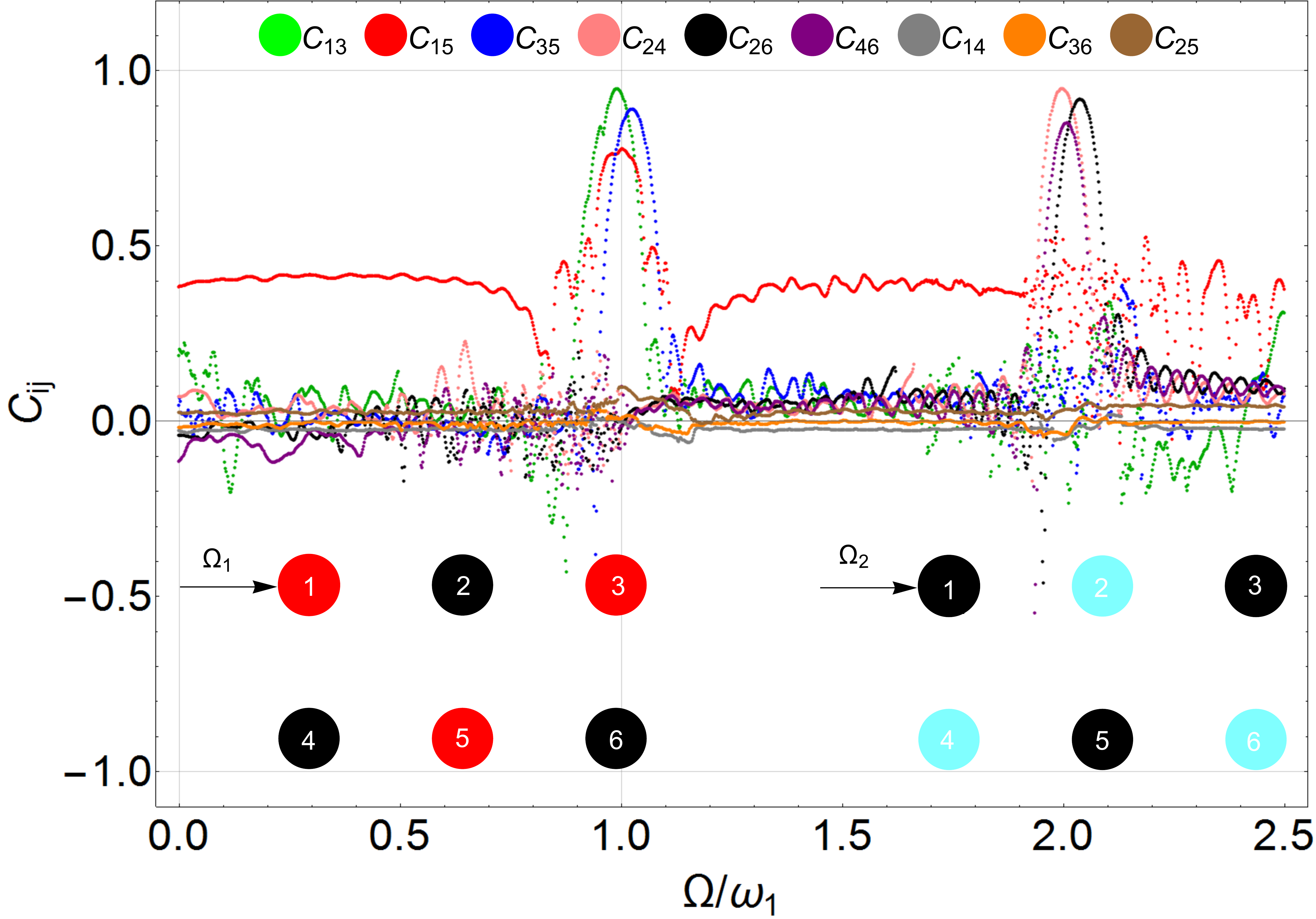}\newline
\caption{(color online) The resonant synchronization of $N=6$ KOs estimated by
$C_{ij}$ versus the driving frequency $\Omega$. The eigenfrequencies of the
oscillators are $\omega_{1}=0.94$, $\omega_{3}=1$, $\omega_{5}=1.06$,
$\omega_{2}=1.97$, $\omega_{4}=2$, $\protect\omega_{6}=2.03$,
respectively, and the coupling strength and the driving strength $ k_{ij}=0.36$, $\Lambda$=1.
Two different driving frequencies are $\Omega_{1}=1$ and $\Omega_{2}=2$. Insets: the networks
of $6$ KOs driven by external signals with frequencies of $\Omega_1$ (left) and $\Omega_2$ (right),
different colors stand for different synchronization groups.}
\label{FIG4}
\end{figure}

In order to show the details of the resonant synchronization, we consider $N=6$ KOs
and analyze the synchronization dynamics among these coupled KOs. In this small network,
we initially set the eigenfrequencies of oscillators $1$, $3$, $5$ randomly around 1.0,
and $2$, $4$, $6$ around 2.0 with a small deviation (see the parameters in Fig.\ref{FIG4}).
Here, the eigenfrequencies set in advance stand for the stored information in the neural
molecular which is characterized by its firing eigenfrequency $\omega_i$. When an external
signal with a driving frequency of $\Omega_{1}=1$ is added to one KO node of the network (see the bottom
left inset of Fig.\ref{FIG4}), the oscillators $1, 3$ and $5$ synchronize into a group with the enhanced
mutual Pearson correlation coefficients shown in Fig.\ref{FIG4} (see the resonant peaks of curves
$C_{13}$, $C_{35}$ and $C_{15}$). If the driving frequency of $\Omega_{2}=2$ is added (see the bottom
right inset of Fig.\ref{FIG4}), the KOs of $2, 4$ and $6$ are synchronized group instead (see the resonant
peaks of curves $C_{24}$, $C_{46}$ and $C_{26}$) due to the resonant synchronization selected by the
external stimuli with different resonant frequencies.

\subsection{The pattern retrieved from the memorized lattice network by resonant synchronization}

\begin{figure}[b]
\includegraphics[width=0.45\textwidth]{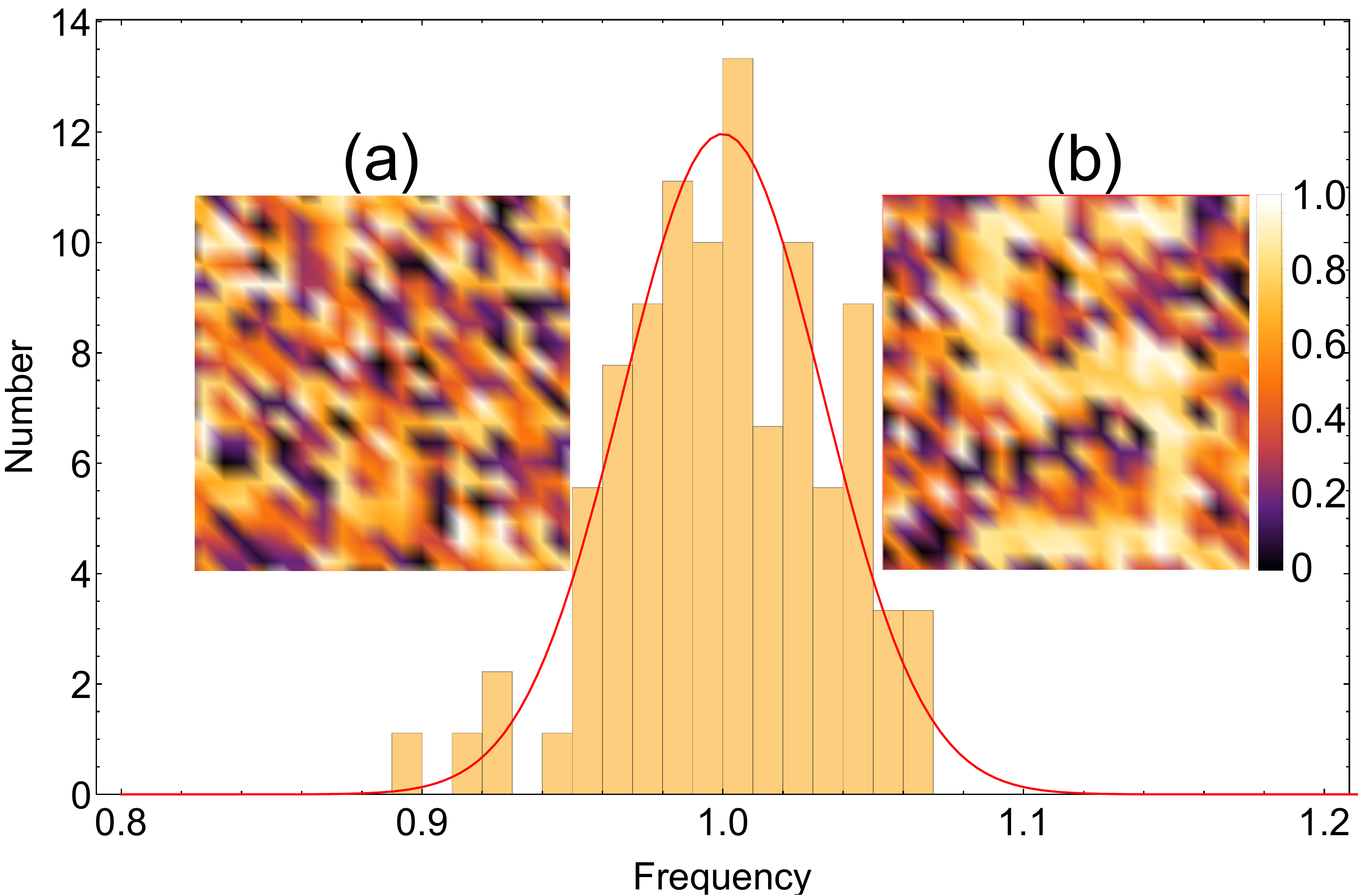}\\
\includegraphics[width=0.45\textwidth]{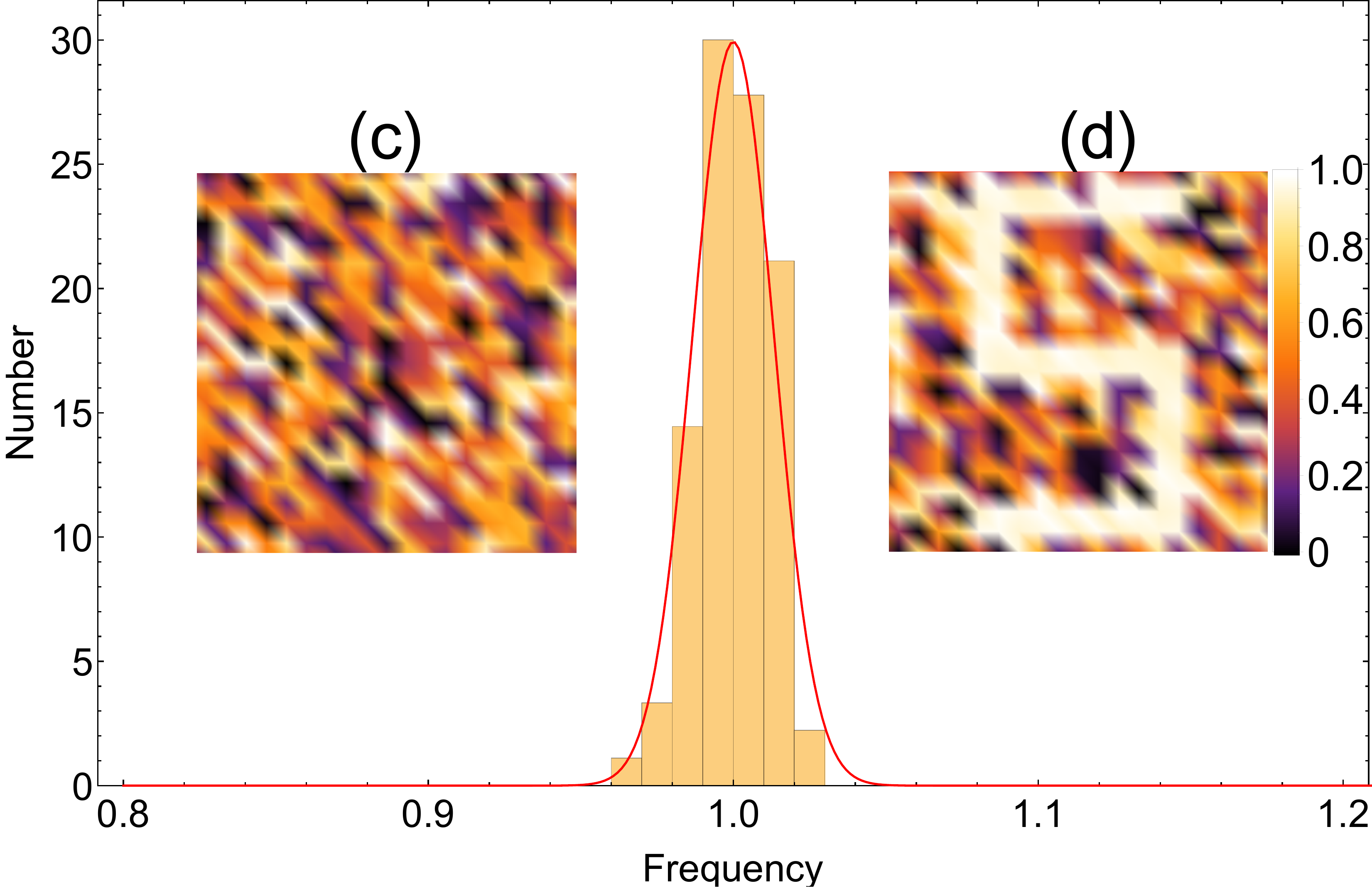}
\caption{(color online). Memory synchronization pattern retrieved from $C_{ij}$ with resonant
driving (b)(d) and without external driving (a)(c). The frequencies of the memorized
oscillators are assigned by Gaussian distributions with mean value of $1$ and the deviations
0.1/3 (the upper one) and 0.04/3 (the lower one), the frequencies of the other
oscillators are set randomly from the interval [0, 2]. The other parameters are $\Lambda=1 , k_{ij}=0.36$.}
\label{FIG5}
\end{figure}

Now, we will show how to recover a stored image in our brain by using well-connected network of
KOs based on the resonant synchronization. As we know, the information in our brain, no matter stored
or retrieved, are all activated by receiving signals inside or outside our mind. Therefore, the memory
is established by the dynamic responses of neutrons to different input stimuli.
The different stimuli will construct different neuron molecules whose properties are characterized by
different dynamic responses such as the firing eigenfrequencies we choose here in our present model.
In order to reveal this, we investigate the synchronized behaviors of neurons in a regular neural network induced by
the external driving sources based on the above modified Kuramoto model.
We use a globally coupled neural network of $20\times 20$ KO lattice to simulate the
well-connected neurons in a local area of cerebral cortex in our brain. Here,
we use frequency to code the information in the neural network. The eigenfrequency
values of the coded KOs are set in advance by a Gaussian distribution with the same mean
value to construct an image of ``5'' through a spatial configuration in the lattice network,
and the eigenfrequencies of the other KOs are assigned randomly (no code area). If the lattice network is
subjected to an external stimulus with a resonant driving frequency, the neurons with closed
eigenfrequencies are activated and then synchronized together by the resonant driving. All the
KOs with enhanced $C_{ij}$ recover a synchronized pattern of $``5"$ emerging in the brain.
The robust effect of resonant synchronization to retrieve the pattern ``5" is shown in Fig.\ref{FIG5}.
The images in the left column, Fig.\ref{FIG5}(a)(c), are displayed by the mutual synchronous
matrix of $C_{ij}$ without adding stimulus signals, while the images in the right column,
Fig.\ref{FIG5}(b)(d), are $C_{ij}$ with stimulus of $\Omega=1$. When we add the driving signal
to the network, the resonant synchronization pattern appears that the synchronized KOs with
similar stored eigenfrequencies are activated and connected with each other by the driving stimulus (right column).
If there is no input, all KOs are oscillating at their own eigenfrequencies independently, and the
dynamics of the whole network is indeed random and no regular pattern is produced (left column).
Therefore, by a direct simulation, we can find a clear pattern emergence of a stored image
if an external stimulus is added.

Considering that the human memory is a complex network with a large number of randomly arranged
neurons immersed in a noisy background, the dynamics of neurons in the human brain may not as
simple as that we considered above, but this simple model can hold the main idea of the searching mechanism
of our memory. When we set the coded KOs' eigenfrequencies with Gaussian distributions of
different deviations in order to consider different intensities
of unavoidable noises in the memory, the resonant synchronous pattern of $C_{ij}$ can be recovered in different resolutions.
A larger deviation will lead to a more fuzzy picture of synchronized
pattern as in Fig.\ref{FIG5}(b) than that of a small deviation one shown in Fig.\ref{FIG5}(d). This phenomenon is
very similar to the memory of our human brain that a lager disturbance will decrease the accuracy of memory
and fails an effective information extraction. Therefore, the memory needs to be repeated by decreasing the
deviation of the information stored in the neuron molecules.
Furthermore, if the neural network is stimulated by different types of stimulus, different synchronized
patterns corresponding to different frequencies of stimuli will also be retrieved. Different information can
be stored in the neurons characterized by different frequency channels and the storage capacity of information
in a network is determined by the frequency discrimination of the network. If the stimuli
are similar, similar mixing patterns will be retrieved form our memory, which is another important characteristic
of the neural network to enable imaginations of our brain.

\section{Conclusion}

In this article, we found a robust cooperative behavior of a globally coupled network, the resonant synchronization, emerged
from the modified Kuramoto oscillators, and can be used to decipher the mechanism of memory searching in the human brain.
Physically, the stimuli from our sensory organs will construct neural networks in our brain with different
characteristics of neurons (derived from different molecular components), whose reactive properties here are denoted by their responsive frequencies of electrical
pulsing. Different frequencies denote different kind of information stored in the neural network. We simulate
the collective responses of the well-connected network to the external stimuli, and the resonant synchronized
pattern of the KOs with similar eigenfrequencies appears in a regular KOs network. The results show that there
exists a significant enhancement of collective dynamics with the synchronized motions induced by external
driving and which exhibits dramatically different behaviors from that without driving.
Under the influence of input driving on a critical network, all the oscillators (nodes) whose frequencies are close to the frequency of external
driving are activated and synchronized, just like the recall process in our memory to get all the resonant (similar)
information by certain stimulus. There are many different types of memories in our brain, the light, the sound or the pressure, etc.,
they all can be selected and extracted by differen types of electric stimuli from our different sensory organs, and the irrelevant
information will be suppressed and can not be activated by resonant effect. The resonant synchronization to retrieve information from our brain is a
quick, efficient and robust information searching way and, basically, it is also a simple and universal phenomenon in physics.
\newline

\begin{acknowledgments}
This work was supported by the National Natural Science Foundation of China (Grant No. 11447025) and the
Scientific Research Foundation for Returned Overseas Chinese Scholars of the State Education Ministry.
\end{acknowledgments}

\end{document}